\documentclass{an}

\usepackage{graphicx,fancyhdr}
\usepackage{times}

\begin{document}

\sloppy
\newcommand{\kms}{km\,s$^{-1}$}
\newcommand{\Halpha}{H$\alpha$}

% -------------- Title and affiliation
%                (please edit accordingly)

\title{Mining the CFHT Legacy Survey for known Near Earth Asteroids}

      \author{O. Vaduvescu\inst{1,2,3,8}\fnmsep\thanks{\email{ovidiuv@ing.iac.es}\newline}, 
              A. Tudorica\inst{4,5,6,7},
              M. Birlan\inst{2}, 
              R. Toma\inst{6,8}, 
              M. Badea\inst{4}, 
              D. Dumitru\inst{4,9}, 
              C. Opriseanu\inst{9}, 
              D. Vidican\inst{9}
              }

\titlerunning{Mining the CFHTLS for known NEAs}
\authorrunning{O. Vaduvescu et al}

   \institute{Isaac Newton Group of Telescopes, 
              Apartado de Correos 321, E-38700 Santa Cruz de la Palma, Canary Islands, Spain \\
            \and
              IMCCE, Observatoire de Paris, 
              77 Avenue Denfert-Rochereau, 75014 Paris Cedex, France \\
            \and
              Instituto de Astronom\'ia, Universidad Cat\'olica del Norte, Avenida
              Angamos 0610, Antofagasta, Chile\\
            \and
              University of Bucharest, Department of Physics, 
              Platforma Magurele, Str. Fizicienilor nr. 1, CP Mg - 11, Bucharest Magurele 76900, Romania \\
            \and
              The Institute for Space Sciences (ISS), 
              Bucharest - Magurele, Ro-077125 Romania \\
            \and
              Rheinische-Friedrich-Wilhelms Universitaet Bonn, Argelander-Institut fur Astronomie, 
              Auf dem Hugel 71 D-53121 Bonn, Germany \\
            \and
              Bonn Cologne Graduate School of Physics and Astronomy, Germany \\
            \and
              The Romanian Society for Meteors and Astronomy (SARM), 
              CP 14 OP 1, 130170, Targoviste, Romania \\
            \and
              Bucharest Astroclub, 
              B-dul Lascar Catargiu 21, sect 1, Bucharest, Romania \\
             }

\date{Received; accepted; published on-line}

%\keywords{astrometry, minor planets, archives, data mining}

% -------------- Abstract

\keywords{asteroids, ephemerides, methods: data analysis, astronomical databases: miscellaneous}

\abstract {Abstract: The Canada-France-Hawaii Legacy Survey (CFHTLS) comprising about 25,000 
           MegaCam images was data mined to search for serendipitous encounters of known Near 
           Earth Asteroids (NEAs) and Potentially Hazardous Asteroids (PHAs). A total of 143 
           asteroids (109 NEAs and 34 PHAs) were found on 508 candidate images which were field 
           corrected and measured carefully, and their astrometry was reported to Minor Planet 
           Centre. Both recoveries and precoveries (apparitions before discovery) were reported, 
           including data for 27 precovered asteroids (20 NEAs and 7 PHAs) and 116 recovered 
           asteroids (89 NEAs and 27 PHAs). Our data prolonged arcs for 41 orbits at first or 
           last opposition, refined 35 orbits by fitting data taken at one new opposition, 
           recovered 6 NEAs at their second opposition and allowed us to ameliorate most orbits 
           and their Minimal Orbital Intersection Distance (MOID), an important parameter to 
           monitor for potential Earth impact hazard in the future. } 

\maketitle

% -----------------------------------------------------------------

\section{Introduction}
\label{intro}

Despite the continuous grow of the existing imaging archives and surveys taken with 
various telescopes around the globe, extremely little work has been devoted to data 
mining in order to ameliorate the orbits of known asteroids and Near Earth Asteroids 
(NEAs) and Potentially Hazardous Asteroids (PHAs). 

Searching for known minor bodies in old imaging archives is not a new idea, such 
work being carried in the last two decades by a few authors in order to recover some 
asteroids and comets and improve their orbits (\cite{bow92}; \cite{hav92}; \cite{mcn95}; 
\cite{bf00}; etc). During the last decade, some dedicated data mining work has been 
carried out to search for known NEAs in a few entire photographic plate archives, 
namely the projects AANEAS (\cite{ste98}, who introduced the term ``precovery''), 
ANEOPP (\cite{boa01}) and DANEOPS (\cite{hah02}). Recently, we presented the public 
server PRECOVERY devoted to search {\it all} known asteroids (including NEAs and PHAs 
besides all other catalogued asteroids) in {\it any} archive uploaded by the user, 
given by a simple observing log recorded in a standard format (\cite{vad09}). 

More than 7,400 NEAs are known today (Nov 2010) and some 1,170 of these are 
catalogued as PHAs according to the JPL NEO database (\cite{nas10}). Many of these 
bodies have been insufficiently observed, i.e. only during about one month of 
visibility at their first opposition. Some of these are classified as Virtual 
Impactors (VIs), while about 70 are considered lost due to their present very 
large uncertainty in their orbits and ephemeris and their very faint brightness, 
according to the NEODyS database (\cite{mil10a}). Based on the currently available 
observations, JPL Sentry System (\cite{nas10}) monitors more than 300 NEAs possibly 
to cause future Earth impact events during the next 100 years, although virtually 
all have almost zero probability to cause such impacts. 

Thanks to five dedicated US-lead surveys searching for NEAs during the last two 
decades, we have discovered the tip of the iceberg of the entire NEA population 
consisting mostly in $\sim1$km and larger objects detectable with 1m class telescopes. 
Nevertheless, sub-km sized asteroids as small as 150m could still cause regional or 
global scale disaster in the eventuality of a catastrophic event (\cite{mor06}), 
thus a common effort should be pursued not only to discover but also to recover and 
follow-up known NEAs. 

During the past 4 years, part of the EURONEAR project we have observed about 200 
selected NEAs using 10 non-dedicated 1-2m telescopes during about 50 nights obtained 
mostly through regular time allocation competition which has been difficult to obtain 
in the absence of a dedicated facility (\cite{bir10}). Besides new observations, data 
mining of existing imaging archives represents another goal of the EURONEAR program, 
and a first paper introduced the method and software to perform the search of any 
archive for known NEAs, PHAs and other asteroids (\cite{vad09}). 

In the present paper we will use the same method to data mine the entire 
Canada-France-Hawaii Legacy Survey (more than 25,000 wide field MegaCam images) 
for known Near Earth Asteroids. Section~\ref{CFHTLS} briefly introduce the survey 
and present the data mining method. Section~\ref{results} will present the results 
grouped in five special classes, and Section~\ref{future} will conclude the paper, 
introducing two related projects in development. 
%__________________________________________________________________

\section{Data Mining of the CFHTLS}
\label{CFHTLS}

\subsection{CFHT Legacy Survey}

Mounted at the prime focus of 3.6m Canada-France-Hawaii Telescope (CFHT) atop Mauna Kea 
in Hawaii, the wide-field imager MegaCam mosaic camera was dedicated in 2003 to become the 
largest field ($\sim1$ square degree) facility available worldwide until 2007 when the 
1.8m Pan-STARRS survey opened, although this facility is still in engineering phase. MegaCam 
consists in 36 CCDs $2048\times4612$ pixel each (340 Mega-pixels total) having a resolution of 
$0.187\arcsec$/pix and producing a total field of view of $0.96\deg \times 0.94\deg$. 

Canada and France joined a large fraction ($\sim50$\%) of their dark and grey telescope time 
from mid-2003 to early 2009 for a large project, the CFHT Legacy Survey (CFHTLS). The data 
acquisition and calibration of the CFHTLS has been a major undertaking for the Canadian and 
French communities, with more than 450 nights over 5 years being devoted to this project by 
CFHT. Based on the diverse science interests of the large CFHT community, CFHTLS includes 
three components: 

\begin{itemize}
\item
The Very Wide survey observed shallow in 3 colours, covering a band of +/-2 degrees along 
the ecliptic for a total area of 410 square degrees, counting 5,980 images; 
\item
The Wide survey observed deeper in 5 colours, covering 170 square degrees in four patches, 
counting 7,295 images; 
\item
The Supernova and Deep survey (very deep and covering only 4 fields observed in 5 
filters at many epochs, counting 12,289 MegaCam images. 
\end{itemize}

At the CFHT User's meeting which took place in 2007 in Marseilles we presented the 
opportunity to search the CFHTLS archives for known NEAs, PHAs and other asteroids 
(\cite{vad07}). In that work we searched the ``candidate images'' of the CFHTLS 
Very Wide component (the most interesting to produce most encounters of asteroids) to find 
serendipitous detections of NEAs, PHAs and all other known asteroids. Both recovery and 
``precovery'' (apparitions before discovery date) were searched using a PHP script which 
queried the SkyBoT server (\cite{imc10}) and the CFHTLS observing log database available 
at the CFHT website (\cite{cfh10}). Overall for the CFHTLS Very Wide component alone, we 
predicted about 450 candidate images probable to hold precovery and recovery apparitions 
of NEAs and PHAs, while an average of 10 known Main Belt asteroids are visible in every 
observed CFHTLS field! 

To search the candidate fields for predicted encounters and measure all such findings, 
we have joined in a team of eight people including five amateur astronomers and students 
and two professional astronomers, so this work is an example of a collaboration between 
professional and amateur astronomers. We present next the necessary steps to perform the 
entire work. 

\subsection{Searching for NEAs using PRECOVERY}

To search for possible serendipitous encounters of all known NEAs, PHAs and other asteroids in 
the CFHTLS archive, we used PRECOVERY, a software written in PHP to perform searches and classify 
findings in any archive (\cite{vad09}). PRECOVERY uses an observing log holding the following basic 
information to define observations: the archive image identifier, observing date (calendar date and
start UT time), telescope pointing ($\alpha$, $\delta$) at J2000.0 epoch, exposure time (sec), image 
field (degrees) and eventually other information. Besides this input file, the software uses the 
asteroid orbital elements database downloaded daily from the Minor Planet Centre (MPC), holding all 
known NEAs, PHAs, numbered and un-numbered asteroids. A dedicated option was built to search the 
CFHTLS/MegaCam archive and is available on site, taking into account the raw format of the CFHTLS 
archive and the geometry of the MegaCam (the position of each of the 36 CCDs forming the entire mosaic). 
For the search, we used the MPC asteroid database of 9 April 2009, thus the CFHTLS archive could 
produce new findings based on a new search to include the asteroids discovered after that date. 

The three CFHTLS survey components add together a total of 25,564 images to search, a slow task 
for one user approach to transfer lots of data and queries between two servers (PRECOVERY and 
SkyBoT). Thus, we divided the big master archive log in batches of 250 images each, which were 
then run individually by the members of the team, one batch at a time by one person, during one 
session with PRECOVERY. Distributed between all members of the team, the search of the entire 
CFHTLS archive took some 20 days to run total time\footnote{Following this work, SkyBoT server 
improved significantly its speed by adding new hardware and an improved search method, so the 
entire job is expected to take much less now.}. 

All candidate images were included in a table listing all data necessary for inspection and 
measurement: the image number and the CCD number, the encountered asteroid name, expected position 
($\alpha$, $\delta$) and its associate uncertainty (in arcsec), expected $V$ magnitude, observing date 
(start of exposure: calendar and Julian date), exposure time (sec) and filter. After applying 
a limiting magnitude $V=24$ (compatible and safer than the survey specifications), we assembled 
these data in a master candidate images database holding about 1,000 images in total to be analysed 
in the next step.

\subsection{Inspection of the Candidate Images}

The master candidate images database was split between members of the team who downloaded
from the Canadian Astronomical Data Centre (CADC) the processed MegaCam detrended images (already 
corrected by overscan, bias, mask and flat-fielding). Using the DS9 (to open MegaCam data cube) 
or IRAF\footnote{IRAF is distributed by the National Optical Astronomy Observatories, which are 
operated by the Association of Universities for Research in Astronomy, Inc., under cooperative 
agreement with the National Science Foundation} (to cut the appropriate image), we split the 
individual corresponding CCD predicted to hold NEAs. Then we used DS9 to inspect visually each 
candidate CCD image close to the predicted ($\alpha$, $\delta$) position, taking into 
account the positional uncertainty of the objects. The inspection task was easily performed by 
blinking subsequent images of the same field, usually found to hold the same object at different 
positions which was easily spotted to move between frames. If only one or two predicted 
images were available to hold a given asteroid, then we downloaded another image closed in time 
of the same field in the same filter to serve for the blinking process, in order to reject 
potentially mis-identification (other asteroids, image flaws, supernovae, galaxies, etc).

\subsection{Uncertain Identification from Few Observations} 
\label{fewobs}

The mining of the CFHTLS archive showed us how important the survey cadence and the search 
work-flow are, also revealing several limitations and possible failures of the precovery work. 
One of the major problems could appear for the objects which could be poorly identified only 
based on very few available images, defined as one or two apparitions only. The problem becomes 
even more difficult in case of poor detection (low S/N due to faint magnitude or/and poor 
weather). Moreover, the situation becomes critical for objects previously observed only at 
one opposition (a few weeks or months only), especially at one epoch very distant in time 
(a few years) from the available observational arc. In this last case, due to the relatively 
poorly determined orbit, the ephemeris uncertainty grows with time and it could reach from 
a few dozen arcsec to a few degrees, thus finding the object becomes much more difficult. 
Obviously, some major question rise in general related to uncertain identifications, namely 
how confident could be these detections and how often these situations arise? 

The detailed answers to the above question are out of the scope of our paper and they depend 
on the survey strategy (number of visits, cadence, exposure time, surveyed area, etc) and also 
on the available statistics for the known NEA population at a given time. Here, we enumerate a 
few criteria to be taken into account for the correct identification in case of few observations,
based on our CFHTLS data mining experience. 

\begin{enumerate}
\item 
Location - In the first step, the search should start closely and around the position 
predicted by some very accurate ephemeris (SkyBoT in this case), taking into account 
the line of variation and the confidence ellipse assumed by some (usually linear) orbital 
uncertainty model (e.g., \cite{mg09}); 
\item 
Apparent Motion - This represents probably the most important criterion for the 
correct identification of a searched object. The observed apparent motion could be assessed 
only from multi-apparitions (if available usually on neighbour images), the predicted motion 
in both $\alpha$ and $\delta$ directions and the time interval between successive exposures. 
The motion information fails in case that only one image is available, and in this case 
other factors should be taken into account; 
\item 
Magnitude - Another important identification criterion is the expected apparent 
magnitude of the object, but two factors usually impede the correct assessment of the 
magnitude, namely the longer exposures (resulting in long trails) and the mostly unknown 
spectral class of the object and its colour (in order to reduce the correct magnitudes); 
\item 
Aspect - Especially when sparse data is available, a very important identification factor 
is the expected aspect of the searched object. Taking into account the pixel scale, exposure 
time, magnitude and the apparent movement of the searched object, its aspect could appear 
either as a long trail (linear, with a thickness compatible with stellar FWHM), a small 
trail (compatible with a slightly elliptical PSF, in which case the trail orientation is 
a very important criterion to be compared with the expected movement direction) or a 
point-like ``stellar'' object (in which case the identification should take into account 
other criteria). If possible, the inspection of the field must be combined with other 
deep-sky images in order to avoid confusions with background galaxies.
\end{enumerate}

Due to the CFHTLS cadence (at least 4 visits of each field from which at least 3 taken in the 
same night), most of our present work involved multi-apparition objects (measurable each on at 
least 3 positions), which total 99 objects (about $70\%$ of the total number of apparitions). 
In these cases, we inspected the fields visually (aligning the images centred on the central 
predicted position, then using the blink) so that the object recognition was obvious, taken into 
account the expected proper motion and magnitude from at least 3 apparitions. The rest of 44 
objects represent few observations objects, with 20 objects appearing only in one image 
($14\%$ of total) and 22 objects appearing in two images (15\%). For all these cases we applied 
the above search criteria, so that only 2 objects could not be measured (about $1\%$ from the 
total reported) due to their faint magnitude resulting in a very high risk of bad detection. 
In case of very faint objects (closed to the limit of CFHTLS detection, about $V\sim23$) or 
poor S/N due to bad weather conditions, we applied {\it boxcar} in IRAF binning $2\times2$ 
in order to improve the S/N for an easier detection. 

\subsection{Field Correction and Measurement}

The detrended CADC images do not include astrometric field correction of the original MegaCam 
images. Field correction is necessary to fix the optical distortion of any wide field camera 
which reaches up to $\sim2\arcsec$ towards the margin of the raw MegaCam field. To remain 
compatible with our past EURONEAR astrometric accuracy ($\sim0.2\arcsec$) and also to take 
full advantage of MegaCam's capability, we had to correct the raw detrended images for the 
field distortion effect. In this sense, we used the available software written at TERAPIX Data 
Centre, specially built to correct MegaCam images and reduce CFHTLS data. 

The field correction process and semi-automatic measurement of the asteroid positions 
consists in five steps, given a CADC CCD distorted field image with header astrometric 
coefficients in the USNO catalogue system. First, we applied SWARP to correct the field 
distortion using the same USNO astrometric system. Second, we used SExtractor to extract 
sources with USNO positions. Third, we applied SCAMP to correct the astrometry from USNO 
to UCAC catalogue system (known to have better accuracy than USNO). Fourth, we used MISSFITS 
to update the header of the corrected image to include UCAC astrometric coefficients. 
Finally, we used again SExtractor to extract all sources from the corrected field 
image in the UCAC reference system. Among all extracted sources (mostly stars and galaxies), 
the coordinates ($\alpha$, $\delta$) of the searched asteroid were extracted from the final 
catalogue. 

Most encounters were found in the Wide field (ecliptic) component for which most exposures 
were small, thus most asteroids appear stellar-like or slightly elliptical, possible to measure 
automatically by SExtractor. Some exposures took longer (e.g., those coming from the two deeper 
surveys) and some asteroids moved faster close to opposition, thus some encounters resulted in 
trails necessary to be measured visually in DS9. We checked all SExtractor findings by inspecting 
the final resulted catalogues overlaid on the DS9 final corrected images. For the bad SExtractor 
findings, we either visually measured the centres of the trails (for the shorter ones) or we 
calculated the centre of the trails by averaging the two ends. Finally, we recorded all 
measured positions together with the observational data in our asteroid master catalogue.

%______________________________________________________________

\section{Results}
\label{results}

\begin{figure}
\centering
\includegraphics[angle=0,width=8cm]{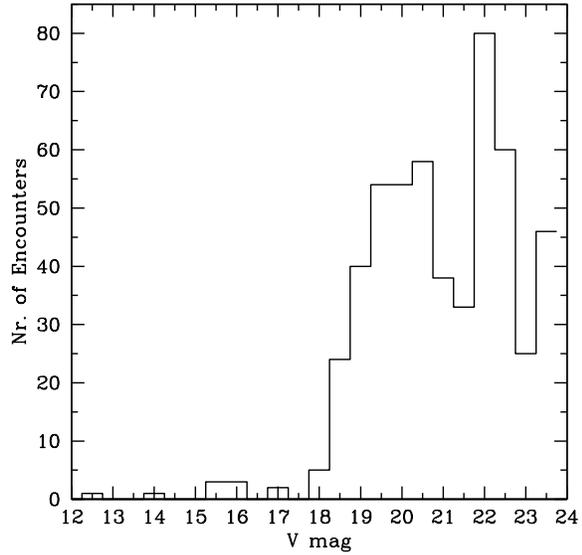}
\begin{center}
\caption{Histogram showing our total number of asteroid encounters (NEAs and PHAs). 
Two main bulks are visible at $V\sim20$ and $V\sim22$ and are discussed in the text. 
}
\label{fig1}
\end{center}
\end{figure}

We encountered 508 candidate images holding a total of 143 NEAs and PHAs whose positions 
($\alpha$, $\delta$) were measured and reported to Minor Planet Centre (MPC). From these, we 
found 109 NEAs (20 NEAs precovered and 89 NEAs recovered) and 34 PHAs (7 PHAs precovered and 
27 PHAs recovered). In average, each asteroid was measured on 3.5 images, which is consistent 
with the Very Wide component which holds most encounters. In the Very Wide component alone 
we found a total of 111 NEAs and PHAs ($78\%$ from total number), in the Wide component 33 
NEAs and PHAs ($22\%$ which confirms that NEAs should be searched all over the sky) and in 
the Supernova and Deep survey none. 

In Figure~\ref{fig1} we plot the histogram showing our total number of encounters (images) 
as a function of the object predicted $V$ magnitude, using a bin of 0.5 mag. Most asteroids 
have magnitudes fainter than $V\rm\sim\rm18$, although a few were found at brighter regime, as 
bright as $V\rm\sim\rm12$. On the plot there are two apparent bulks visible. The first bulk peaks 
around $V\rm\sim\rm20$ and probably represents the objects observed close to their opposition 
accessible to other established 1m surveys. The second bulk peaks around $V\rm\sim\rm22$ and 
correspond to objects inaccessible to the other dedicated surveys, possibly representing 
objects either fainter or not observed at opposition, and in this regime the CFHTLS could 
bring a more important contribution. 

\subsection{Astrometry}

We submitted 508 measured positions to Minor Planet Centre (MPC) and most of them ($99\%$)
were accepted. They were included in the MPC, NEODyS and other databases and they were taken 
into account by major providers for the amelioration of the orbits. Only two objects (positions) 
were rejected by the MPC, to which we will refer in Section~\ref{secondopp}.

\begin{figure}
\centering
\includegraphics[angle=0,width=8cm]{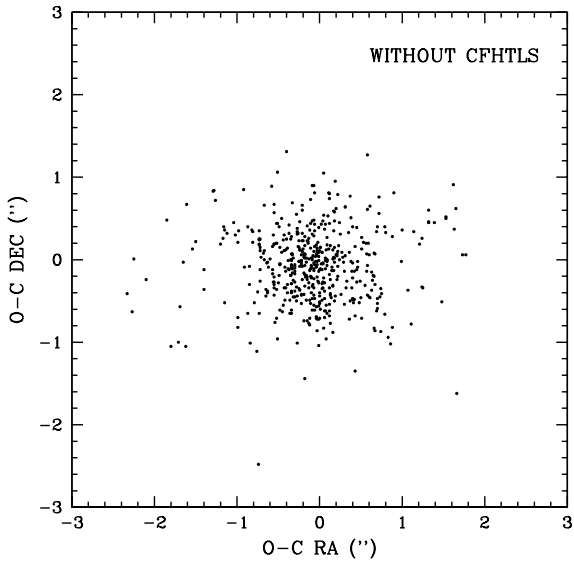}
\begin{center}
\caption{O-C (Observed minus Calculated) residuals in $\alpha$ and $\delta$, 
where the calculated positions refer to orbits which do not include our data. 
The average standard deviation is $0.97\arcsec$ and the sample standard deviation 
is $2.91\arcsec$ and the plot includes all 508 measured positions. A few points 
referring to orbits with larger residuals are outside the limits. 
}
\label{fig2}
\end{center}
\end{figure}

\begin{figure}
\centering
\includegraphics[angle=0,width=8cm]{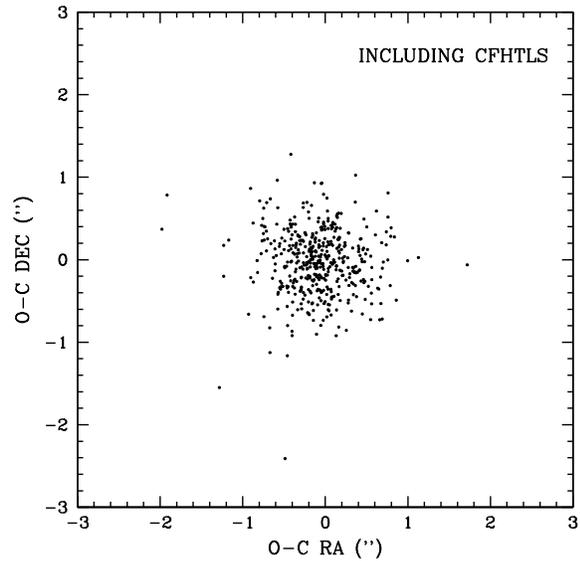}
\begin{center}
\caption{O-C (Observed minus Calculated) residuals in $\alpha$ and $\delta$, 
where the calculated positions refer to orbits which include our data. The average 
standard deviation is $0.24\arcsec$ and the sample standard deviation is $0.38\arcsec$. 
The plot includes all accepted positions and all points are inside the limits. 
}
\label{fig3}
\end{center}
\end{figure}

In Figure~\ref{fig2} we plot the O-C residuals (Observed minus Calculated in $\alpha$ 
and $\delta$) resulting from orbits which do not include our data. One could observe some 
relatively large spread of the residuals around the origin, with a larger spread in the 
right ascension, consistent with the proper motion of most asteroids. The average standard 
deviation is $0.97\arcsec$ and the sample standard deviation is $2.91\arcsec$ and the 
plot includes all 508 measured positions. Some points are located outside the visible 
limits of the plot which focuses on the central region for a better view. 
In Figure~\ref{fig3} we plot the same data resulted from orbits fitted with all available 
data set including our CFHTLS accepted observations. Most points are better confined around 
the origin, dropping the average standard deviation to $0.24\arcsec$ and the sample 
standard deviation to $0.38\arcsec$. The points are better grouped around the centre in 
Figure~\ref{fig3} compared with Figure~\ref{fig2}, probing the fact that the majority 
of the initial orbits could be well adjusted after including our data. Compared with 
previous statistics, the deviations obtained from the orbits which include our data probe 
that our work contributed to the refinement of the orbits. We will discuss these findings 
further. 

\subsection{Amelioration of Orbits}

We evaluated our contribution to the final orbital solution which includes our data. In this
sense, we used NEODyS observational data available to date 10 Jan 2010. From the total of 143 
asteroids found in the survey, 58 orbits resulted to be interesting to be studied ($40\%$), 
and we group them in 5 special classes based on the existing observing data available before 
our data mining. We include these results in Table~\ref{table1}. 
Besides the asteroid name, we include the MPC classification, the number of CFHTLS observations, 
the orbital arc before and after adding our data (where ``w'' stands for weeks, ``m'' for months 
and ``y'' for years), the number of covered oppositions before and after adding our data, 
and some comments showing how our data improved the available orbits. These special cases are 
presented in the next sections. 

We compared the orbits fitted with and without our observations, using all other available 
observations taken from the NEODyS (.rwo) database. To fit orbits, we used the ORBFIT package 
(\cite{mil10b}) running first ORBFIT to fit the available observations using full differential 
corrections and the nonlinear least squares method, then running FITOBS to propagate the orbit 
to the same epoch and perform a close approach analysis which includes an iterative calculation
of MOID (Minimal Orbital Intersection Distance) in 10 steps. A similar comparison using the 
ASTERPRO software (\cite{roc07}) and FIND\_ORB software (\cite{gra10}) produced similar results. 

Table~\ref{table2} includes the most notable cases of orbits ameliorated with our CFHTLS data. 
We calculated with ORBFIT the six Keplerian orbital elements for the epoch $MJD=55400.0$: the 
semi-major axis ($a$), eccentricity ($e$), inclination of the orbit ($i$), longitude of the ascending 
node ($\Omega$), argument of the pericenter ($\omega$) and the mean anomaly ($M$). To assess the 
potential impact hazard and the goodness of the fit we include the calculated MOID, the number of 
fitted observations and the residual mean square RMS of the fit. For each asteroid we give in the 
first line the orbital fit including our data and in the second line the orbital fit excluding 
our observations. 

Comparing the orbital elements, one could observe that most orbits were improved slightly: $a$ 
and $e$ change mostly at the 4-5th decimal (representing up to 15,000 Km in semi-major axis), the 
angles $i$, $\Omega$, $\omega$ change at their 3-4th decimal in most cases, and $M$ changes mostly 
at its 1st or 2-nd decimal. Comparing the goodness of the fit (counted by the RMS in the $\sigma$ 
column), the majority of the orbits improved after including our data, namely $\sigma$ decreased by 
$0.01-0.02\arcsec$ and in some cases up to $0.04\arcsec$ (PHA 1993 BX3) and $0.05\arcsec$ 
(NEA 2004 QE20) - both objects being observed only by us at their last opposition. Looking at 
the MOID column, most of the orbits became less chaotic after fitting our data, converging faster 
in our 10 step iterative process calculated by ORBFIT - the MOID intervals became narrower, e.g. 
for 2003 TG2 of the second set in Table~\ref{table2} the initial MOID obtained without our data 
varies between 0.19460 and 0.19565 AU (an interval 0.00105 AU), while after fitting our data it 
varies between 0.19497 and 0.19551 (an interval 0.00054 AU), so in this case we constrained MOID 
by 0.00051 AU = 76,500 Km. In some cases MOIDs were changing at their 5-th or 4-th decimal 
(representing up to 15,000 Km), although in many cases they remain unchanged. 

We present next the five special classes derived from the existing observing data and orbital arcs 
available before our work. 

\subsubsection{Extended Arcs at First Opposition (Precoveries)}

A total of 21 asteroids (15 NEAs and 6 PHAs) were precovered, i.e. found on 75 images taken before 
their discovery date (as recorded by MPC) and we include them in the first group of Table~\ref{table1}. 
From these, 7 asteroids (5 NEAs and 2 PHAs) have their 1st or 2-nd opposition covered only by the 
present work (reported as X/(X+1) in the ``Opp'' column). For the rest of 14 asteroids (11 NEAs and 
3 PHAs) we have prolonged their arcs with data at first opposition and we give in the Comments 
column the extended interval. We improved the existent orbits and MOIDs by fitting our data to the 
previous observations and this can be observed in the columns $\sigma$ and MOID in the Table~\ref{table2}. 
As one can see comparing the first line (including our observations) versus the second, MOIDs converge 
better while RMS' decrease after including our data for the majority of the objects. 
Six cases deserve to be evidenced based on their extended Arc column: 2008 ED69 (having an orbital 
arc data prolonged from 9 months to 4 years), 2005 OW and 2005 QN11 (having their short arcs prolonged 
by one month), 2008 AF4 (PHA very desirable having a MOID = 0.00281 and the orbital arc extended from 4 
months to 6 years), 2007 FS35 (arc extended from 3 months to 8 years) and 2008 CR118 (with the arc 
extended from 8 months to 5 years). In two other cases we could constraint the MOIDs (2007 RM133 and 
2005 UU3). We compare the fitted orbits in Table~\ref{table2}. 

\subsubsection{Extended Arcs at Last Opposition (Recoveries)}

A total of 14 asteroids (9 NEAs and 5 PHAs) were recovered by us at their last opposition. From 
these, 7 objects (4 NEAs and 3 PHAs) have their last opposition covered only thanks to our work. 
All orbits were improved by fitting our CFHTLS data. 
At least four objects deserve to be noted based on the extending time coverage (given in the Arc 
column): 1998 VD35 (PHA desirable with the orbital arc prolonged with 5 years), 2005 WA1 (PHA 
extremely desirable having the short arc improved from one month to 7 months), 2003 TG2 and 2004 
XG29 (NEAs having the very short arcs of 18 and 25 days prolonged by 6 and 10 days, respectively). 
For two other objects we decreased the RMS, namely for 1993 BX3 (a numbered object) by $0.04\arcsec$ 
and for 2004 QE20 (NEA very desirable) by $0.05\arcsec$. We compare the fitted orbits in 
Table~\ref{table2}. 

\subsubsection{Refined Arcs at one Intercalated Opposition}

A total of 15 asteroids (all NEAs from which 10 are considered desirable or very desirable) 
were recovered by us in the CFHTLS at one intercalated opposition, and our data represent the 
only available observational set at the indicated opposition (given in the Comments column). 
Most objects had observed data at many oppositions, so their orbits could be improved only
marginally, nevertheless one object merits attention, namely 1998 QB28 (NEA very desirable) 
for which we could constrained the MOID by 0.00007 AU, as can be observed in Table~\ref{table2}. 

\subsubsection{Refined Very Small Arcs}

Two NEAs have their very short orbits improved thanks to our work. 2005 YD was observed only 
for two weeks (26 observations) for which we reported two more observations weighting about 7\% 
from the entire data set, while 2008 RZ24 was observed for two months being found by us in the Very 
Wide survey in 11 images which weight for about 17\% of the entire data set. Their orbits could be 
improved using our data, as one can see in the $\sigma$ column in Table~\ref{table2}. 

\subsubsection{Extended Arcs at Second Opposition (Major Recoveries)}
\label{secondopp}

A total of six asteroids (all NEAs extremely desirable) were observed previously only at one
opposition, only for a few months. They were found by us in 13 CFHTLS images within $2\arcsec$ 
to $50\arcsec$ distance from their predicted positions, consistent with their orbital $1\sigma$ 
uncertainty ellipse calculated at their observing date by NEODyS. 
Two of them (2006 UD17 and 2007 VX137) appeared only on one image each, so we dropped them 
due to high risk of mis-identification from noise. These were the only objects rejected by MPC 
(\cite{spa10}). The other four objects appear on multiple (2 or 3) images and they had systematic 
O-C residuals. Their apparitions verified most of the criteria presented in Section~\ref{fewobs}, 
so we reported them to MPC who accepted the data. 
These notable cases of major recoveries are: 2005 OJ3 (precovered by us 2 years before its 
discovery in 2005), 2008 CJ70 (precovered 3 years before discovery), 2000 SZ44 (recovered by 
us 5 years after its oldest discovery in 2000) and 2002 VR94 (recovered by us after 2.5 years 
following discovery in 2002). All orbits and MOIDs could be improved with our data for all
reported objects, as one could check in Table~\ref{table2}. 

%______________________________________________________________

\section{Future Work}
\label{future}

We continue to offer PRECOVERY to the community for other data mining projects 
(\cite{vad10}), and our server will offer soon new focused search capabilities. 
Recently we have proposed two similar projects to expand our data mining work. 

\subsection{The Archives ESO/WFI and INT/WFC}

In a team of about 10 people including mostly students and amateur astronomers, 
in autumn 2009 we have embarked in a project to data mine the 2.2m ESO/MPG Wide 
Field Imager archive and the INT 2.5m Wide Field Camera archive. These comprise of 
about 100,000 and 230,000 images respectively, taken during the last decade by two 
similar wide field ($34\arcmin\times34\arcmin$) cameras mounted on similar 2m class 
telescopes located in both hemispheres. This project is about half completed, and 
has already built the two databases from the off-line nightly observing logs of 
the ING and on-line ESO Data Archive for the ESO/MPG. We have run PRECOVERY on both 
these archives, inspecting about 1,500 ESO candidate images and measuring a few 
hundred positions, and this project continues. 

\subsection{MEGA-PRECOVERY and the Mega-Archive}

Recently we have started to write a code (named MEGA-PRECOVERY) to address a new 
data mining method focused on a list of a few specified known objects (NEAs or PHAs) 
to search a ``mega-archive'' comprising in a number of given archives whose observing 
logs will be available soon on the EURONEAR website. To start this ``mega-archive'',
we will join the CFHTLS, ESO/WFC, INT/WFI and Bucharest plate archives, and we plan 
to add soon the DSS, SDSS, and later the Wide Field Plate Database (WFPDB, 
www.skyarchive.org) which stores the archive pointings of about one thousand plate 
archives existing worldwide (\cite{tsv91,tsv05}). 
Empowered by this new tool to data mine this proposed ``mega-archive'', we plan to 
propose to international forums such as IVOA and IAU to ask every observatory to make 
available in a first phase their observing logs in a standard VO format to be data mined 
for any poorly known asteroid. Besides the available existing data, we consider that the 
continuous exponential grow due to recent and new surveys could make such a data 
mining tool more than rewarding, and we consider that our present paper proved this. 
%______________________________________________________________

\begin{acknowledgements}
This project was based on observations obtained with MegaPrime/MegaCam, a joint project of 
CFHT and CEA/DAPNIA at the Canada-France-Hawaii Telescope (CFHT) which is operated by  the 
National Research Council (NRC) of Canada, the Institut National des Science de l'Univers 
of the Centre National de la Recherche Scientifique (CNRS) of France, and the University of 
Hawaii. This work is based in part on data products produced at TERAPIX and the Canadian 
Astronomy Data Centre (CADC) as part of the Canada-France-Hawaii Telescope Legacy Survey, 
a collaborative project of NRC and CNRS. 
Alin Nedelcu and Thierry Lim helped with the measurement of some raw CFHTLS images which 
allowed us to compare raw data with data corrected by SWARP suite, and Patrick Rocher kindly 
provided us his ASTERPRO code to compare with ORBFIT and FIND\_ORB. 
Thanks are due to ORBFIT consortium for sharing their code of the ORBFIT package. 
Acknowledgements are also due to Bill Gray, the author of the FIND\_ORB software, for his 
very prompt assistance in order to install and run his code which we found very flexible 
and better than ORBFIT in the fitting process of initial orbits. 
We are also endowed to Emmanuel Bertin for making his SExtractor suite available to correct 
and measure the CFHTLS detreneded images, and also to Jerome Berthier for his continuous 
support with the SkyBoT server accessed by PRECOVERY. This research has made use of SAOImage 
DS9, developed by Smithsonian Astrophysical Observatory. 
This research has made use of IMCCE's SkyBoT VO tool (\cite{ber06}). We are thankful 
to Minor Planet Centre, specifically to Tim Spahr and Brian Marsden who pointed out our 
initial errors in the reported positions. Dr. Tim Spahr also served as the referee of our 
paper and his comments helped us to improve its content. 

\end{acknowledgements}

\appendix
\section{Appendix: Data Tables}

\renewcommand{\arraystretch}{0.8}
\begin{table*}[!t]
\begin{center}	
\caption{Five special classes including 58 NEA and PHA asteroids data mined in the CFHTLS. 
Besides the asteroid name we give its MPC classification, the number of CFHTLS observations, 
the orbital arc and the number of covered oppositions before and after adding our data, and 
some comments showing how our work improved the orbits. }
\label{table1}
\begin{tabular}{llrrll}
\hline
\hline
\noalign{\smallskip}
\noalign{\smallskip}
Asteroid  &  Classification  &  Obs  &   Arc  &   Opp  &  Comments \\
\noalign{\smallskip}

\hline
\noalign{\smallskip}\noalign{\smallskip}
\multicolumn{6}{c}{Extended Arcs at First Opposition (Precoveries):} \\
\noalign{\smallskip}
 2008 ED69  &  NEA very desirable     &  6  &  9m/4y & 2/3 &  Arc prolonged by 3 yrs    \\
 2005 CJ    &  PHA very desirable     &  3  &  5/8m  &  2  &  Arc prolonged by 3 mths   \\
 2006 PA1   &  PHA very desirable     &  1  &  4y    &  3  &  Arc prolonged by one month  \\
 2008 OX2   &  PHA                    &  4  &  2y    &  2  &  Arc prolonged by 1.5 mths \\
 2003 WO151 &  NEA very desirable     &  3  &  2y    &  2  &  Arc prolonged by 1.5 mths \\
 2005 LW    &  NEA very desirable     &  2  &  4/5y  & 3/4 &  Arc prolonged by 8 mths   \\
 2005 OW    & NEA extremely desirable &  3  &  4/5m  &  1  &  Short arc prolonged by 1 mth \\
 2005 QN11  & NEA extremely desirable &  3  &  4/5m  &  1  &  Short arc prolonged by 1 mth \\
 2005 QS10  &  NEA very desirable     &  3  &  4y    &  2  &  Arc prolonged by 1.5 mths \\
 2005 SS4   &  NEA very desirable     &  4  &  3y    &  3  &  Arc prolonged by 2 weeks \\
 2004 BE86  &  NEA very desirable     &  4  &  5y    &  2  &  Arc prolonged by one month \\
 2007 RM133 &  NEA                    &  8  &  3y    &  2  &  Arc prolonged by one week  \\
 2008 SQ1   &  NEA                    &  5  &  5y    &  2  &  Arc prolonged by one month \\
 2008 AF4   &  PHA very desirable     &  1  &  4m/6y & 2/3 &  We only at 2nd opp, Goldstone radar target \\
 2007 FS35  &  NEA very desirable     &  4  &  3m/8y & 2/3 &  We only at 2nd opp \\
 2008 CR118 &  PHA                    &  1  &  8m/5y & 2/3 &  We only at 2nd opp \\
 2006 SV19  &  NEA                    &  3  &  6y    & 3/4 &  We only at 2nd opp, numbered (212546) \\
 2006 SU49  &  PHA very desirable     &  3  &  7y    & 3/4 &  We only at 2nd opp  \\
 2005 RN33  &  NEA very desirable     &  6  &  4y    &  2  &  We first at 2nd opp \\
 2008 XE3   &  NEA                    &  4  &  4y    &  2  &  We 2nd set at 1st opp \\
 2005 UU3   &  NEA very desirable     &  4  &  2y    &  2  &  We 2nd set, only just 4 hrs after discovery \\
\noalign{\smallskip}

\hline
\noalign{\smallskip}\noalign{\smallskip}
\multicolumn{6}{c}{Extended Arcs at Last Opposition (Recoveries):} \\
\noalign{\smallskip}
 1998 VD35  &  PHA desirable          &  1  &  2/7y  &  3/4  &  Arc prolonged by 5 yrs, numbered (20425) \\
 1993 BX3   &  PHA desirable          &  6  & 11/13y &  3/4  &  Arc prolonged by 5 yrs, numbered (65717) \\
 1999 GS6   &  PHA desirable          &  3  &  7/8y  &  4/5  &  Arc prolonged by 1 yr, numbered (152754) \\
 2005 RR6   &  PHA very desirable     &  4  &  2y    &  2    &  Arc prolonged by 2 weeks \\
 2005 WA1   & PHA extremely desirable &  3  &  1/7m  &  1    &  Initial 3 week arc prolonged by 6 months \\
 2003 TG2   & NEA for survey recovery &  3  & 18/24d &  1    &  Very small arc prolonged by one week, old object \\
 2004 XG29  & NEA extremely desirable &  1  & 25/35d &  1    &  Very small arc prolonged by 10 days \\
 1998 XA5   &  NEA very desirable     &  3  &  4/8y  &  3/4  &  Arc prolonged by 4 yrs \\
 2002 TY57  &  NEA very desirable     &  1  &  3/5y  &  2/3  &  Arc prolonged by 2 yrs \\
 2002 AA    &  NEA very desirable     &  6  &  5y    &  3    &  Arc prolonged by 1 week \\
 2007 DL8   &  NEA very desirable     &  4  &  2y    &  2    &  Arc prolonged by 2 mths \\
 2003 TX9   &  NEA very desirable     &  6  &  3y    &  2    &  Arc prolonged by 6 mths \\
 2002 AC29  &  NEA very desirable     &  3  &  7y    &  3/4  &  Arc prolonged by 3 mths \\
 2004 QE20  &  NEA very desirable     &  3  &   5/7y & 3/4   &  Arc prolonged by 2 yrs, numbered (164221) \\
\noalign{\smallskip}

\hline
\noalign{\smallskip}\noalign{\smallskip}
\multicolumn{6}{c}{Refined Arcs at one Intercalated Opposition:} \\
\noalign{\smallskip}
 2001 OE84  &  NEA                    &  2  &  8y    &  3/4  &  We alone at 3rd opp \\
 1997 GH3   &  NEA desirable          & 10  & 13y    &  4/5  &  We alone at 3rd opp, numbered (19356)  \\
 2002 LS32  &  NEA very desirable     &  1  &  8y    &  5/6  &  We alone at 4th opp \\
 1998 QB28  &  NEA very desirable     &  3  &  9y    &  2/3  &  We alone at 2nd opp \\
 1999 RP36  &  NEA                    &  3  & 10y    &  3/4  &  We alone at 3th opp, numbered (217683) \\
 2003 CJ11  &  NEA desirable          &  4  &  4y    &  3/4  &  We alone at 2nd opp, numbered (154453) \\
 1998 ST4   &  NEA very desirable     &  3  & 11y    &  5/6  &  We alone at 4th opp \\
 2000 YM29  &  NEA                    &  4  & 10y    &	5/6  &  We alone at 4th opp, numbered (153219) \\
 2002 TS67  &  NEA very desirable     &  2  &  8y    &  3/4  &  We alone at 2nd opp \\
 2005 WS55  &  NEA                    &  3  &  7y    &  3/4  &  We alone at 2nd opp, numbered (209924) \\
 2008 LW8   &  NEA very desirable     &  2  & 12y    &  3/4  &  We alone at 3rd opp \\
 2000 DH8   &  NEA                    &  4  & 15y    &  5/6  &  We alone at 3rd opp, numbered (231792) \\
 1993 TQ2   &  NEA very desirable     &  5  & 15y    &  2/3  &  We alone at 2nd opp \\
 2000 UP30  &  NEA very desirable     &  8  &  7y    &  2/3  &  We alone at 2nd opp \\
 2001 WL15  &  NEA very desirable     &  4  &  9y    &  4/5  &  We alone at 4th opp \\
\noalign{\smallskip}

\hline
\noalign{\smallskip}\noalign{\smallskip}
\multicolumn{6}{c}{Refined Very Small Arcs:} \\
\noalign{\smallskip}
 2005 YD    & NEA for survey recovery &  2  &  2w    &  1    &  \\
 2008 RZ24  & NEA extremely desirable & 11  &  2m    &  1    &  \\
\noalign{\smallskip}

\hline
\noalign{\smallskip}\noalign{\smallskip}
\multicolumn{6}{c}{Extended Arcs at Second Opposition (Major Recoveries):} \\
\noalign{\smallskip}
 2005 OJ3   & NEA extremely desirable &  3  &  8m/2y & 1/2   &  O-C=$6\arcsec$  V=23.6, precovery 2 yrs before discovery \\
 2008 CJ70  & NEA extremely desirable &  2  &  3m/3y & 1/2   &  O-C=$50\arcsec$ V=23.1, precovery 3 yrs before discovery \\
 2000 SZ44  & NEA extremely desirable &  3  &  4m/5y & 1/2   &  O-C=$13\arcsec$ V=22.4, recovery 5 yrs after discovery \\
 2002 VR94  & NEA extremely desirable &  3  &  6m/3y & 1/2   &  O-C=$30\arcsec$ V=23.6, recovery 2.5 yrs after discovery \\
\noalign{\smallskip}

\hline
\hline
\end{tabular}
\end{center}
\end{table*}

% ________________

\begin{table*}[!t]
\begin{center}
\caption{Comparison of the orbits fitted with (first line) and without our observations (second line). 
         Keplerian orbital elements fitted with ORBFIT at epoch $MJD=55400.0$: the asteroid name, semimajor axis $a$, 
         eccentricity $e$, inclination $i$, longitude of the ascending node $\Omega$, argument of pericenter $\omega$ 
         and mean anomaly $M$, followed by the minimal orbital intersection distance MOID, number of fitted observations 
         and the squared mean residual RMS of the fit. } 
\label{table2}
\resizebox{17.3cm}{!}{
\begin{tabular}{lrrrrrrrrr}
\hline
\hline
\noalign{\smallskip}
\noalign{\smallskip}
Asteroid  & $a$ (AU) & $e$ & $i$ ($\deg$) & $\Omega$ ($\deg$) & $\omega$ ($\deg$) & $M$ ($\deg$) & MOID (AU) &  Obs & $\sigma$ ($\arcsec$)\\
\noalign{\smallskip}

\hline
\noalign{\smallskip}\noalign{\smallskip}
\multicolumn{10}{c}{Extended Arcs at First Opposition (Precoveries):} \\
\noalign{\smallskip}

2008 ED69  &  2.88704287 &  0.74949654 &  36.27922752 &  149.89327262 &  172.73282884 &  149.61802749 &  0.28316 &  116 &  0.43	\\ 
           &  2.88695213 &  0.74948772 &  36.27908600 &  149.89317168 &  172.73280795 &  149.62504553 &  0.28316 &  110 &  0.42	\\ 
\noalign{\smallskip}
2005 OW    &  2.66552267 &  0.60163695 &   1.63921135 &  271.76312432 &   62.27499199 &   46.31259102 &  0.05759 &  196 &  0.62  \\ 
           &  2.66553757 &  0.60163914 &   1.63921642 &  271.76315442 &   62.27495821 &   46.30915905 &  0.05758 &  193 &  0.63  \\ 
\noalign{\smallskip}
2005 QN11  &  2.17394532 &  0.40379176 &  5.61935281 &  223.87836246 &  134.99008565 &  184.49056673 &  0.30336    &  121 &  0.46 \\ 
           &  2.17393231 &  0.40378871 &  5.61933855 &  223.87834461 &  134.99036525 &  184.49533346 &  0.30330-38 &  118 &  0.46 \\ 
\noalign{\smallskip}
2007 RM133 &  2.21037753 &  0.44000603 &  10.74595065 &  106.19581007 &  181.01826765 &  347.88273273 &  0.22113-18 &  56 &  0.51 \\ 
           &  2.21036767 &  0.44000347 &  10.74591063 &  106.19601557 &  181.01822181 &  347.88492253 &  0.22112-19 &  48 &  0.53 \\ 
\noalign{\smallskip}
2008 AF4   &  1.38256104 &  0.41072419 &  8.91934131 &  109.42271956 &  293.32280895 &  231.52478785 &  0.00281 &  609 &  0.35	\\ 
           &  1.38256494 &  0.41072640 &  8.91938330 &  109.42273385 &  293.32278690 &  231.52224732 &  0.00281 &  606 &  0.35	\\ 
\noalign{\smallskip}
2007 FS35  &  1.92227709 &  0.39022490 &   0.31760987 &  183.27038985 &  107.04010819 &   31.13370375 &  0.15568-71 &  60 &  0.45 \\ 
           &  1.92238624 &  0.39026668 &   0.31758960 &  183.26936559 &  107.03613297 &   31.10509908 &  0.15565-72 &  53 &  0.43 \\ 
\noalign{\smallskip}
2008 CR118 &  1.83875731 &  0.51066465 &  3.92343947 &  121.63512526 &  156.91147019 &  286.31967420 &  0.02816   &  81 &  0.44 \\ 
           &  1.83879655 &  0.51067494 &  3.92353435 &  121.63581110 &  156.91013530 &  286.31072755 &  0.02815-6 &  74 &  0.43 \\ 
\noalign{\smallskip}
2005 UU3   &  1.28261561 &  0.47819728 &  13.93810052 &  36.53446656 &  128.56296260 &  27.45528293 &  0.14251-303 &  44 &  0.41 \\ 
           &  1.28263495 &  0.47820262 &  13.93815480 &  36.53441590 &  128.56421885 &  27.45518844 &  0.14250-306 &  40 &  0.41 \\ 
\noalign{\smallskip}

\hline
\noalign{\smallskip}\noalign{\smallskip}
\multicolumn{10}{c}{Extended Arcs at Last Opposition (Recoveries):} \\
\noalign{\smallskip}
1998 VD35  &  1.56459680 &  0.47673984 &  6.98207379 &  227.41633118 &  296.12600123 &  294.06492318 &  0.00321 &  51 &  0.58 \\ 
           &  1.56459674 &  0.47673982 &  6.98209499 &  227.41637403 &  296.12599271 &  294.06499563 &  0.00321 &  50 &  0.58 \\ 
\noalign{\smallskip}
1993 BX3   &  1.39463215 &  0.28060259 &  2.79020747 &  175.58505195 &  289.94925112 &  233.79801622 &  0.04843 &  53 &  0.74	\\ 
           &  1.39463214 &  0.28060257 &  2.79020832 &  175.58505307 &  289.94925038 &  233.79802668 &  0.04843 &  47 &  0.78	\\ 
\noalign{\smallskip}
2005 WA1   &  2.00712579 &  0.58526544 &  10.93346025 &  247.38964489 &  241.55518760 &  212.65535613 &  0.02070 &  118 &  0.62	\\ 
           &  2.01068769 &  0.58610164 &  10.94631051 &  247.39020632 &  241.55363801 &  211.12777389 &  0.02049-88 &  115 &  0.62 \\ 
\noalign{\smallskip}
2003 TG2   &  0.90787297 &  0.31598894 &  25.44938968 &  200.70288030 &  355.13821055 &  109.12183940 &  0.19497-551 &  35 &  0.58 \\ 
           &  0.90782816 &  0.31593624 &  25.43375232 &  200.70922357 &  355.13053872 &  109.34478265 &  0.19460-565 &  32 &  0.59 \\ 
\noalign{\smallskip}
2004 XG29  &  1.40962299 &  0.31319954 &  0.15454852 &  302.84078467 &  109.89518303 &  141.58385674 &  0.00205 &  130 &  0.73 \\ 
           &  1.40960282 &  0.31318696 &  0.15454391 &  302.83963116 &  109.89759837 &  141.63226231 &  0.00205 &  129 &  0.74 \\ 
\noalign{\smallskip}
2004 QE20  &  1.50507593 &  0.20534407 &  6.48274424 &  272.66090730 &  74.16056708 &  67.14803071 &  0.22006 &  130 &  0.57 \\ 
           &  1.50507608 &  0.20534454 &  6.48272684 &  272.66089645 &  74.16057837 &  67.14785564 &  0.22006 &  129 &  0.62 \\ 
\noalign{\smallskip}

\hline
\noalign{\smallskip}\noalign{\smallskip}
\multicolumn{10}{c}{Refined Arcs at One Intercalated Opposition:} \\
\noalign{\smallskip}
1998 QB28  &  2.07448980 &  0.37976447 &  1.07717741 &  341.64613291 &  297.97833026 &  23.54298808 &  0.27085-47 &  42 &  0.37	\\ 
           &  2.07448933 &  0.37974343 &  1.07719064 &  341.64645854 &  297.97695080 &  23.54609416 &  0.27093-48 &  39 &  0.34	\\ 
\noalign{\smallskip}

\hline
\noalign{\smallskip}\noalign{\smallskip}
\multicolumn{10}{c}{Refined Very Small Arcs:} \\
\noalign{\smallskip}
2005 YD    &  1.65283640 &  0.42520216 &  4.78305977 &  90.65373286 &  314.23849158 &  73.57127069 &  0.02261 &  28 &  0.52 \\ 
           &  1.65241013 &  0.42504894 &  4.78167561 &  90.65428349 &  314.23914129 &  73.87912964 &  0.02260 &  26 &  0.53 \\ 
\noalign{\smallskip}
2008 RZ24  &  2.17784533 &  0.56163149 &  13.93533754 &  165.75988542 &  122.24834247 &  228.94219608 &  0.07838-40 &  63 &  0.41 \\ 
           &  2.17787177 &  0.56163720 &  13.93543395 &  165.75996916 &  122.24824768 &  228.93803259 &  0.07838-40 &  52 &  0.43 \\ 
\noalign{\smallskip}

\hline
\noalign{\smallskip}\noalign{\smallskip}
\multicolumn{10}{c}{Extended Arcs at Second Opposition (Major Recoveries):} \\
\noalign{\smallskip}
2005 OJ3   &  2.71013672 &  0.53762893 &  4.44043486 &  239.00829721 &  154.97116812 &  11.35563295 &  0.26280-1 &  65 &  0.45 \\ 
           &  2.71021194 &  0.53764106 &  4.44045440 &  239.00783784 &  154.97121905 &  11.34024651 &  0.26280-2 &  62 &  0.46 \\ 
\noalign{\smallskip}
2008 CJ70  &  1.40566635 &  0.15171298 &  17.33745102 &  145.71702268 &  69.89141510 &  109.41544705 &  0.28369-70 &  75 &  0.54 \\ 
           &  1.40568600 &  0.15171773 &  17.33805493 &  145.71693753 &  69.89010832 &  109.40607845 &  0.28367-74 &  73 &  0.54 \\ 
\noalign{\smallskip}
2000 SZ44  &  2.44314896 &  0.50419701 &  5.69470263 &  128.83931580 &  250.57411766 &  203.53577984 &  0.23621-22 &  46 &  0.76 \\ 
           &  2.44313484 &  0.50419447 &  5.69468409 &  128.83919426 &  250.57433239 &  203.54361778 &  0.23618-25 &  43 &  0.78 \\ 
\noalign{\smallskip}
2002 VR94  &  2.38103120 &  0.55880939 &  5.57530694 &  57.06182464 &  326.87070923 &  37.18697396 &  0.07397 &  107 &  0.58 \\ 
          &  2.38096936 &  0.55879728 &  5.57523792 &  57.06197398 &  326.87061638 &  37.21641152 &  0.07397-6 &  104 &  0.58 \\ 
\noalign{\smallskip}

\hline
\hline
\end{tabular}
}
\end{center}
\end{table*}


\begin{thebibliography}{}
   \bibitem[Bertier et al 2006]{ber06} Bertier J. et al: 2006, {\it SkyBoT, a new VO service to identify Solar System objects}, 
                   Astronomical Data Analysis Software and Systems XV, 351, 367
   \bibitem[Birlan et al 2010]{bir10} Birlan M. et al: 2010, {\it More than 160 near Earth asteroids observed in the EURONEAR network}, 
                          Astronomy \& Astrophysics, 511, 40
   \bibitem[Boattini and Forti 2000]{bf00} Boattini, A. and Forti, G.: 2000, Planetary and Space Science, 48, 939
   \bibitem[Boattini et al. 2001]{boa01} Boattini, A., et al.: 2001, Astronomy \& Astrophysics, 375, 293
   \bibitem[Bowell 1992]{bow92} Bowell, E.: 1992, IAU Circ. 5585 and 5586
   \bibitem[CFHT 2010]{cfh10} Canada-France-Hawaii Telescope (CFHT): 2010, {\it Full ASCII catalog of exposures per CFHTLS survey}, 
                          http://www.cfht.hawaii.edu/Science/CFHTLS-DATA/exposureslogs.html
   \bibitem[Gray 2010]{gra10} Gray, B.: 2010, {\it FIND\_ORB Software Package}, http://www.projectpluto.com/find\_orb.htm
   \bibitem[Hahn, 2002]{hah02} Hahn, G.: 2002, {\it DLR-Archenhold Near Earth Objects Precovery Survey (DANEOPS)}, 
                          http://earn.dlr.de/daneops/
   \bibitem[Haver et al. 1992]{hav92} Haver, R., et al.: 1992, IAU Circ. 5670
   \bibitem[IMCCE 2010]{imc10} Institute de Mecanique Celeste et de Calcul des Ephemerides (IMCCE): 2010, SkyBoT server, 
                          http://vo.imcce.fr/webservices/skybot/
   \bibitem[McNaught 1995]{mcn95} McNaught, R. H.: 1995, IAU Circ. 6198
   \bibitem[Milani and Gronchi 2009]{mg09} Milani, A. and Gronchi, G.: 2009, {\it Theory of Orbit Determination}, Cambridge Univ. Press, ISBN-13: 9780521873895
   \bibitem[Milani et al 2010a]{mil10a} Milani, A. et al.: 2010a, {\it Near Earth Objects Dinamic Site}, http://unicorn.eis.uva.es/neodys
   \bibitem[Milani et al 2010b]{mil10b} Milani, A. et al.: 2010b, {\it ORBFIT Software Package}, http://adams.dm.unipi.it/~orbmaint/orbfit/
   \bibitem[Morrison 2006]{mor06} Morrison, D: 2006, {\it Asteroid and comet impacts: the ultimate environmental catastrophe}, 
                          Phil. Trans. R. Soc. A, 364, 2041
   \bibitem[NASA 2010]{nas10} NASA: 2010, {\it Near Earth Object Program}, http://neo.jpl.nasa.gov
   \bibitem[Rocher 2007]{roc07} Rocher, P.: 2007, {\it Notes scientifiques et techniques de l'IMCCE}, ISBN: 2-910015-54-8
   \bibitem[Spahr 2010]{spa10} Spahr T.: 2010, private communication
   \bibitem[Steel et al 1998]{ste98} Steel, D., et al.: 1998 Australian Journal of Astronomy, 7, 67
   \bibitem[Tsvetkov 1991]{tsv91} Tsvetkov M. K., IAU Commission 9 Working Group on Wide-Field Imaging, 
                        Newsletter No. 1, p. 17. http://www.skyarchive.org/wgss\_newsletter/issue1/wfpa.pdf
   \bibitem[Tsvetkov 2005]{tsv05} Tsvetkov, M. K.: 2005, {\it Plate Content Digitisation, Archive Mining and Image Sequence Processing},   
                           Astro workshop, Sofia, 2005, ISBN-10 954-580-190-5, p. 10-41.
   \bibitem[Vaduvescu and Curelaru 2007]{vad07} Vaduvescu, O., Curelaru, L.: 2007, {\it Mining NEAs and Asteroids in the CFHTLS}, 
                          CFHT Users Meeting, 9-11 May 2007, Marseille
   \bibitem[Vaduvescu et al 2009]{vad09} Vaduvescu, O. et al: 2009, {\it EURONEAR: Data mining of asteroids and Near Earth Asteroids}, 
                          Astronomische Nachrichten, 330, 7, 698
   \bibitem[Vaduvescu et al 2010]{vad10} Vaduvescu, O. et al: 2010, {\it EURONEAR website: Observing Tools - Archive Precovery and Recovery}, 
                          http://euronear.imcce.fr

\end{thebibliography}
\end{document}